\documentclass[letterpaper, preprint, paper,11pt]{AAS}	% for preprint proceedings

\usepackage{bm}
\usepackage{amsmath}
\usepackage{subfigure}
\usepackage[colorlinks=true, pdfstartview=FitV, linkcolor=black, citecolor= black, urlcolor= black]{hyperref}
\usepackage{overcite}
\usepackage{footnpag}			      	% make footnote symbols restart on each page
\usepackage[export]{adjustbox} 
\usepackage{tabularx} % Required for tabularx environment
\usepackage{arydshln} % Required for \arrulerule
\usepackage{dblfloatfix}
\usepackage{booktabs}
\usepackage{placeins}

\PaperNumber{24-407}

\begin{document}

\title{Orbit Determination through Cosmic Microwave Background Radiation}

\author{Pedro K. de Albuquerque\thanks{Research Affiliate, C5I Center, George Mason University, 4400 University Drive, Research Building, Fairfax, VA, 22030-4444, USA.},  
Andre R. Kuroswiski\thanks{Research Affiliate, Department of Computer Science, University of Central Florida, Orlando, FL 32816-2362, USA},
Annie S. Wu \thanks{Associate Professor, Department of Computer Science, University of Central Florida, Orlando, FL 32816-2362, USA},
Willer G. dos Santos\thanks{Professor, Aeronautics Division, Instituto Tecnologico de Aeronautica, Praça Marechal Eduardo Gomes 50, São José dos Campos, SP, 12228-900, Brazil.},
\ and Paulo Costa\thanks{Director, C5I Center, George Mason University, 4400 University Drive, Research Building, Fairfax, VA, 22030-4444, USA.}
}

\maketitle{}

\begin{abstract}
This research explores the use of Cosmic Microwave Background (CMB) radiation as a reference signal for Initial Orbit Determination (IOD). By leveraging the unique properties of CMB, this study introduces a novel method for estimating spacecraft velocity and position with minimal reliance on pre-existing environmental data, offering significant advantages for space missions independent of Earth-specific conditions. Using Machine Learning (ML) regression models, this approach demonstrates the capability to determine velocity from CMB signals and subsequently determine the satellite's position. The results indicate that CMB has the potential to enhance the autonomy and flexibility of spacecraft operations.
\end{abstract}

\section{Introduction}

Initial Orbit Determination (IOD) is crucial for successfully executing spacecraft missions \cite{raol1985orbit}. The ability to initially ascertain a spacecraft's position, velocity, or orbital elements at a given moment lays the foundation for precise maneuvers, course corrections, and the reliable operation of both navigation systems and onboard payloads.

The trend towards autonomy within the spacecraft sector has accelerated in response to the increasing number of deployed satellites and the growing complexity of missions \cite{starek2015spacecraft}. This shift is particularly vital when spacecraft experience malfunctions that require system resets. In such instances, the capability for a spacecraft to autonomously calibrate its position becomes indispensable to ensuring mission objectives can continue without interruption.

Moreover, while traditional IOD methods relying on ground-based sensors have proven effective \cite{vallado2001fundamentals}, they necessitate that satellite orbits are carefully orchestrated to remain visible to these sensors \cite{beckman2003orbit}. This requirement sometimes conflicts with the ideal trajectories needed to meet mission objectives \cite{mcinroy1995optimal}. Although expanding ground sensor coverage through establishing new sites is a viable solution, it incurs significant financial and logistic challenges.

The evolving landscape of space exploration underscores the critical need for innovative IOD strategies that bolster operational flexibility and lessen the reliance on terrestrial infrastructure. This shift is vital for fostering more robust and autonomous spacecraft missions. While current autonomous IOD methods have made substantial strides by utilizing fixed points such as celestial bodies \cite{beckman2003orbit}, landmarks \cite{christian2019autonomous}, or planetary limbs \cite{christian2017accurate} and occasionally integrating magnetic field data \cite{Psiaki}, they often hinge on specific preexisting environmental knowledge. For example, employing the measurement of the direction for celestial bodies necessitates previous data on the ephemerides of these bodies \cite{Battin1999}. Similarly, navigating around a small body such as an asteroid requires initial estimations of its geophysical characteristics, typically obtained through extensive observation campaigns and refined through in-flight data collection, albeit with high computational costs and inherent limitations \cite{panicucci2021autonomous}.

Another sensor that can be used for IOD is the Global Positioning System (GPS) sensor \cite{haines2003initial}. GPS is a highly accurate and widely used option for satellite navigation in Earth's orbit, providing essential positioning and timing information \cite{yoon2000spacecraft,kang2006precise}. It can even be utilized for missions to the Moon with the appropriate precautions and adjustments \cite{basile2015gps}. However, its application to other celestial bodies is not feasible. Therefore, new general methods are necessary for future space exploration.

The proposal to use only velocity measurements for IOD represents a revolutionary shift that can open new trends in autonomous IOD \cite{christian2019initial}. Picking this idea and introducing the Cosmic Microwave Background (CMB) as an innovative reference signal for velocity determination can demand minimal prior information in the activities of IOD. It was shown that the integration of the CMB signal with Celestial Navigation techniques forms a comprehensive navigation solution under a fusion using an Unscented Kalman Filter \cite{albuquerque2024}. This solution is adept at precisely determining a spacecraft's velocity and position, marrying the age-old principles of Celestial Navigation with the avant-garde application of CMB radiation to craft a holistic approach to space navigation.

Despite the CMB's potential to redefine navigation practices, its application in the context of IOD still needs to be explored, which marks a significant void in astrodynamics research. This paper aims to bridge this gap by presenting a method that utilizes CMB radiation for velocity measurement, sidestepping some limitations inherent in current IOD methodologies. By capitalizing on this signal, this study introduces a technique less dependent on previously acquired environmental knowledge, offering a substantial advantage for IOD in space missions that do not consider Earth-specific conditions or artificial signals.

The main contributions of this research are:
\begin{itemize}
    \item Introduces the use of CMB signals for IOD, providing a unique reference for velocity measurement;
    \item Discusses the theoretical basis of using three velocity vectors for IOD, addressing challenges in CMB measurement, and presenting essential equations;
    \item Develops a non-linear system for velocity determination based on CMB, supported by comprehensive equations and assumptions, and examines the impact of orbital variations;
    \item Introduces a Machine Learning (ML) approach for velocity estimation using CMB signals, including the rationale for using artificial intelligence and the training process;
    \item Showcases simulation results for analytical and ML models and discusses the practical implications of the findings.   
\end{itemize}

\section{Background}

Solving the IOD problem using only velocity vectors resembles addressing Gibbs' problem, yet it is essential to understand that they are not identical. Gibbs' problem involves determining the orbit of a body in space given three position vectors and their corresponding times \cite{Battin1999}. It is a classical method that has laid the groundwork for further developments in orbit determination.

The IOD method enhanced by three velocity vectors expands on the existing concept. An initial analytical method to address this variation exists \cite{christian2019initial}, whereas a more sophisticated geometric solution is provided and will be utilized in this research \cite{hollenberg2020geometric}. 

First, it is necessary to find the normal direction $\vec{k}$ for the orbit, considering all $n$ velocity measurements ($\vec{v}_{s p_i}$, with $i$ varying from $1$ to $n$):
\begin{equation}
\label{eq:1}
\left[\begin{array}{c}
\vec{v}_{s p_1} \\
\vec{v}_{s p_2} \\
\vdots \\
\vec{v}_{s p_n}
\end{array}\right] \vec{k}=\vec{0}_{n \times 1}
\end{equation}
where the solution for $\vec{k}$ is linked with the smallest singular value in the SVD decomposition. Subsequently, it is necessary to construct a new orbital frame  $\{\hat{u}_x,\hat{u}_y,\vec{k}\}$ to make an orthographic projection of the velocity vectors onto the orbital plane. The components of the orbital frame are defined as follows:
\begin{equation}
\label{eq:2}
\hat{u}_x=\vec{v}_{s p_1}^T \times \vec{k}
\end{equation}
\begin{equation}
\label{eq:3}
\hat{u}_y=\vec{k} \times \hat{u}_x
\end{equation}

The project velocities $\vec{v}_{O_i}$ are defined as:
\begin{equation}
\label{eq:4}
\vec{v}_{O_i}^T = \left[\begin{array}{c}
\dot{x}_{O_i} \\
\dot{y}_{O_i}
\end{array}\right]=\left[\begin{array}{ccc}
1 & 0 & 0 \\
0 & 1 & 0
\end{array}\right] \left[\begin{array}{c}
\hat{u}_x^T \\
\hat{u}_y^T\\
\vec{k}^T
\end{array}\right] \vec{v}_{s c_i}^T
\end{equation}

The next step is to fit the hodograph circle with center $\{\dot{x}_{c},\dot{y}_{c}\}$ and radius $R$ to the projected velocity vectors in the orbital plane, solving the following linear system:
\begin{equation}
\label{eq:5}
\left[\begin{array}{ccc}
2 \dot{x}_{O_1} & 2 \dot{y}_{O_1} & -1 \\
2 \dot{x}_{O_2} & 2 \dot{y}_{O_2} & -1 \\
\vdots & \vdots & \vdots \\
2 \dot{x}_{O_n} & 2 \dot{y}_{O_n} & -1
\end{array}\right]\left[\begin{array}{c}
\dot{x}_c \\
\dot{y}_c \\
g
\end{array}\right]=\left[\begin{array}{c}
\dot{x}_{O_1}^2+\dot{y}_{O_1}^2 \\
\dot{x}_{O_2}^2+\dot{y}_{O_2}^2 \\
\vdots \\
\dot{x}_{O_n}^2+\dot{y}_{O_n}^2
\end{array}\right]
\end{equation}
where $g$ is an intermediate variable to find the radius as:
\begin{equation}
\label{eq:6}
R=\sqrt{\dot{x}_c^2+\dot{y}_c^2-g}
\end{equation}

Furthermore, the orbit’s eccentricity vector $\vec{e}$ and the center of the circle of the hodograph may be computed as:
\begin{equation}
\label{eq:7}
\vec{e}=\frac{\vec{c}}{R} \times \vec{k}
\end{equation}
\begin{equation}
\label{eq:8}
\vec{c}=\left[\begin{array}{c}
\hat{u}_x^T \\
\hat{u}_y^T\\
\vec{k}^T
\end{array}\right]^T\left[\begin{array}{c}
\dot{x}_c \\
\dot{y}_c \\
0
\end{array}\right]
\end{equation}

Finally, it is necessary to obtain the position vector $\vec{r}_{s p_i}$ represented as:
\begin{equation}
\label{eq:9}
\vec{r}_{s p_i}=\rho_{s p_i}\hat{r}_{s p_i}
\end{equation}
where $\rho_{s p_i}$ is the magnitude, and $\hat{r}_{s p_i}$ is the direction of the velocity at time $i$. However, it is necessary to calculate the velocity component $\vec{v}_{s p \perp_i}$ to obtain this direction. This component resides within the plane orthogonal to the line connecting the planet and the spacecraft. The hodograph's geometry aids in determining the velocity's direction:
\begin{equation}
\label{eq:10}
\hat{r}_{s p_i}= \frac{\vec{v}_{s p_i}^T-\vec{c}}{\|\vec{v}_{s p_i}^T-\vec{c}\|} \times \vec{k}
\end{equation}

Consequently, $\vec{v}_{s p \perp_i}$ can be obtained as:
\begin{equation}
\label{eq:11}
\vec{v}_{s p \perp_i}=\left(\mathbf{I}_{3 \times 3}-\hat{r}_{s p_i}^T\hat{r}_{s p_i}\right) \vec{v}_{s p_i}^T
\end{equation}
which in turn facilitates the calculation of the magnitude of the velocity:
\begin{equation}
\label{eq:12}
\rho_{s p_i}=\frac{\mu}{R\|\vec{v}_{s p \perp_i}\|}
\end{equation}
with $\mu$ being the central body’s gravitational parameter. 

If velocity measurements present noise, there is an approach that can prove more efficient, particularly if multiple velocity measurements exist. This approach allows for an initial orbit estimation that considers all the measurements and then determines positions \cite{Christian2019}. In this research, a simulation of a circular orbit will be used to focus on fundamental aspects, thus simplifying the work and not requiring the more complex noise-handling methods. 

The CMB will be used as a reference signal to complete the velocity measurements in Equation~(\ref{eq:1}). The CMB, known as the echo of the Big Bang, formed shortly after the recombination period \cite{durrer2020cosmic}. Due to the universe's expansion, it now lies within the microwave spectrum, discovered by Penzias and Wilson in 1965 \cite{penzias1965measurement}. Two key characteristics make the CMB an ideal navigational signal: its isotropy \cite{Souradeep2006} and stable black body behavior \cite{smoot1999cobe}.

Isotropy ensures that the signal is uniform in all directions, which is crucial for consistent navigation references across different spatial points. Black body behavior indicates a stable and predictable spectral profile, making the signal an ideal standard for calibrating instruments and systems for navigation due to its reliability and universality.

However, the CMB does exhibit anisotropies that are Gaussian \cite{akrami2020planck}, with the most significant being 1 in 100,000 \cite{smoot1999cobe}. These anisotropies are often analyzed using spherical harmonics, with the dipole component (moment equal to 1) being a candidate for navigation purposes due to its kinematic nature \cite{meerburg}. Other spherical harmonic components excite cosmology, offering insights into the universe's fundamental properties \cite{Challinor_2012}. 

\begin{figure}[h]
	\centering\includegraphics[width=3.5in]{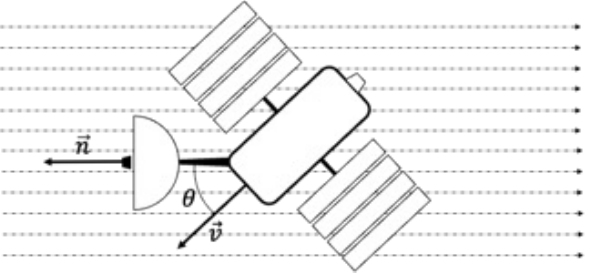}
	\caption{Spacecraft with Velocity Vector $\mathbf{\vec{v}}$ and Sensor Pointing $\vec{n}$ for CMB Detection}
	\label{fig:1}
\end{figure}

Condon and Harwit \cite{Condon1968}, Heer and Kohl \cite{Heer1968}, and Peebles and Wilkinson \cite{Peebles1968}, using transformations of energy and angles, established that if an observer is moving with a velocity $\vec{v}$ relative to a thermal bath (Figure \ref{fig:1}), it perceives radiation differently in each direction. These researchers derived the following equation:
\begin{equation}
\label{eq:13}
T(\vec{n})=\left(\frac{\sqrt{1-|\vec{v} / c|^2}}{1-(\vec{v} / c) \cdot \vec{n}}\right) T_0 = T_D + T_0
\end{equation}
where $T_0$ denotes the isotropic temperature of the CMB, measured to be $2.7255 \pm 0.0009$~K \cite{fixsen1996cosmic}, $T_D$ corresponds to the signal resulting from the observer's movement relative to the radiation field, $c$ represents the speed of light, and $\vec{n}$ symbolizes the direction in which the sensor is oriented. It is observed that, at the given magnitude of $T_0$, a measurement aligned with the velocity vector is anticipated to yield a minuscule temperature variation on the order of $3\times10^{-3}$~K for a velocity of $370$~km/s. This speed corresponds to the relative velocity of the Solar System Barycenter (SSB) with respect to the CMB \cite{Darling2022}.

The feasibility of using CMB for navigation raises two critical issues:  the technological capability to measure the CMB's slight variations due to spacecraft motion and the impact of implementing CMB sensors on spacecraft platforms \cite{albuquerque2024}. Technological advances prove that measuring the CMB's dipole variation from space is feasible, as demonstrated by missions such as COBE \cite{mather1993cosmic}, WMAP \cite{bennett2003microwave}, and Planck \cite{tauber2004planck}. Although the primary interest of some of these missions was to measure higher moments, not only the dipole, across all directions in the sky, the precision of CMB anisotropy measurements has improved dramatically, almost in line with Moore's Law \cite{Abitbol2017CMBS4TB}.

The first issue is related to the capability of measuring slight changes in the CMB. Several factors significantly impact the precision of estimating peculiar velocities from CMB data \cite{notari2012measuring}:
\begin{itemize}
    \item Cosmic variance and instrumental noise limit accuracy, specifically at large multipoles due to cosmic variance and at small multipoles due to instrumental noise;
    \item The instrument's resolution determines the highest multipole that can be reliably measured; and
    \item The sky coverage influences statistical uncertainty, with more coverage reducing this variance.
\end{itemize}

Distinguishing between inherent fluctuations in the CMB and those induced by a spacecraft's velocity is challenging during navigation. However, there are techniques for differentiating these sources, including frequency-specific observations, polarization measurements, advanced signal processing, modeling and calibration, high multipole measurements, and complex scan patterns \cite{albuquerque2024}. Collectively, these methods enhance the accuracy and reliability of the CMB-based sensor system, though some still require further technological development.

For instance, the Planck mission achieved a velocity measurement precision of approximately $60$~km/s using high multipole measurements \cite{notari2012measuring}. In contrast, a future experiment capable of measuring temperature and polarization multipoles up to $5000$ could achieve a precision as fine as $8$~km/s. This illustrates the significant advancements in measurement capabilities that can be expected with the development of more sophisticated technologies and methodologies.

The second issue is more straightforward as the impact varies with the technology employed. CMB sensors, which are generally bolometers or radiometers, require protection from direct sunlight and galactic disc measurements. Additionally, some components need to be maintained at cryogenic temperatures \cite{tauber2004planck}. These requirements could make CMB sensors less appealing despite the advantages of CMB's isotropy. Nevertheless, the isotropic nature of the CMB offers significant navigational benefits. Since the radiation is uniform in all directions, spacecraft equipped with CMB sensors do not need to maintain a specific orientation to gather data, allowing for greater flexibility in their operational configurations. This freedom from having to favor specific pointing directions enables the selection of more optimal orbits, not restricted solely by the need for line-of-sight (LOS) communication with Earth. Consequently, this can significantly enhance a spacecraft's autonomy, allowing it to undertake longer or more complex missions with less dependence on direct Earth-based control and support. These advantages suggest a promising potential for incorporating CMB sensors into future space exploration strategies despite the challenges associated with their implementation.

This work focuses on leveraging the CMB for IOD purposes in space exploration, specifically by measuring radiation intensity transformed into temperature readings by the instruments. However, as already mentioned, these measurements are often contaminated with foreground and systematic noise, which can obscure the subtle temperature fluctuations crucial for precise navigation.

Although the research does not directly address the strategies for mitigating such noise, it is essential to acknowledge that these challenges are assumed to be manageable within the sensor's Time-Ordered Data (TOD). Techniques to clean the signal from these types of noise are well-documented in the literature and considered effective \cite{albuquerque2024,kurki2009destriping,keihanen2004maximum,wallis2017optimal}.

Furthermore, for this research, anisotropies beyond the dipole moment are treated as Gaussian noise \cite{thommesen2020monte}. So, the temperature measured can be presented as follows:
\begin{equation}
\label{eq:14}
T_{CMB}=T_0+T_D+T_F
\end{equation}
In this context, $T_0$ signifies the uniform temperature, $T_D$ corresponds to the signal resulting from the observer's movement relative to the radiation field, and $T_F$ reflects the signal due to density variations in the CMB just prior to the last scattering surface. The sum of $T_0$ and $T_D$ can be summarized as the Equation~(\ref{eq:13}). This assumption simplifies the data analysis, allowing us to concentrate on how the CMB can be harnessed for next-generation sensor technologies in space exploration.

\section{Mathematical Modeling for IOD Using CMB}

To devise a method for velocity determination using CMB sensors within the IOD framework, it is paramount first to understand how variations in certain variables influence temperature readings. When examining Equation~(\ref{eq:13}), it becomes evident that two primary factors can alter the measured temperature: the magnitude of the velocity and the angle between the sensor's pointing and velocity direction.

For the analysis, a hypothetical scenario in which the satellite moves straight in an environment uniformly bathed in thermal radiation from a specific direction is considered (Figure \ref{fig:1}).

\begin{figure}[h!]
	\centering\includegraphics[width=3in]{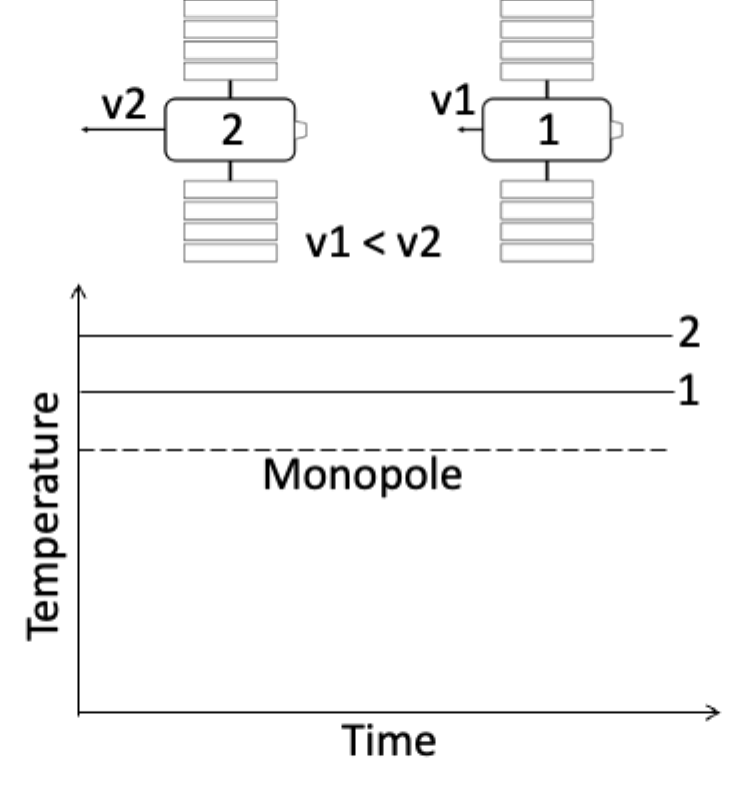}
	\caption{Effect of Varying Velocity Magnitudes ($\mathbf{v_1}$ and $\mathbf{v_2}$) on the Measurement of the CMB Temperature}
	\label{fig:2}
\end{figure}

In such scenarios, an increase in velocity would cause a shift in the measured temperature relative to the monopole baseline, $T_0$. This baseline represents the uniform temperature measured if the observer were stationary with respect to the radiation source. As shown in Figure \ref{fig:2}, spacecraft $2$ is moving at a higher velocity ($v_2$) compared to spacecraft $1$ ($v_1$), so spacecraft $1$ records a temperature closer to the monopole baseline.

\begin{figure}[h]
	\centering\includegraphics[width=3in]{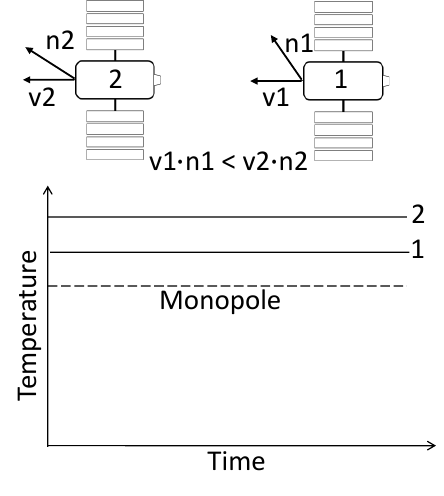}
	\caption{Effect of Sensor Pointing ($\mathbf{n_1}$ and $\mathbf {n_2}$) and Velocity ($\mathbf{v_1}$ and $\mathbf{v_2}$) on the Measurement of the CMB Temperature}
	\label{fig:3}
\end{figure}

The sensor's pointing with the direction of flight also plays a crucial role. A sensor aligned with the flight direction will experience a more significant shift in temperature, with the slightest shift occurring at a 90-degree angle to the direction of flight. In Figure \ref{fig:3}, $n_1$ and $n_2$ represent the sensor pointing, while $n_1 \cdot v_1$ and $n_2 \cdot v_2$ denote the dot products of the velocity and sensor pointing. Spacecraft 2, with $n_2 \cdot v_2$ greater than $n_1 \cdot v_1$, shows a higher temperature displacement to the monopole baseline. Conversely, flight in the opposite direction of the CMB flow will lower the temperature curve, while flight in alignment and in the same direction as the CMB flow will elevate the temperature curve above the monopole baseline.

\begin{figure}[h!]
	\centering\includegraphics[width=3in]{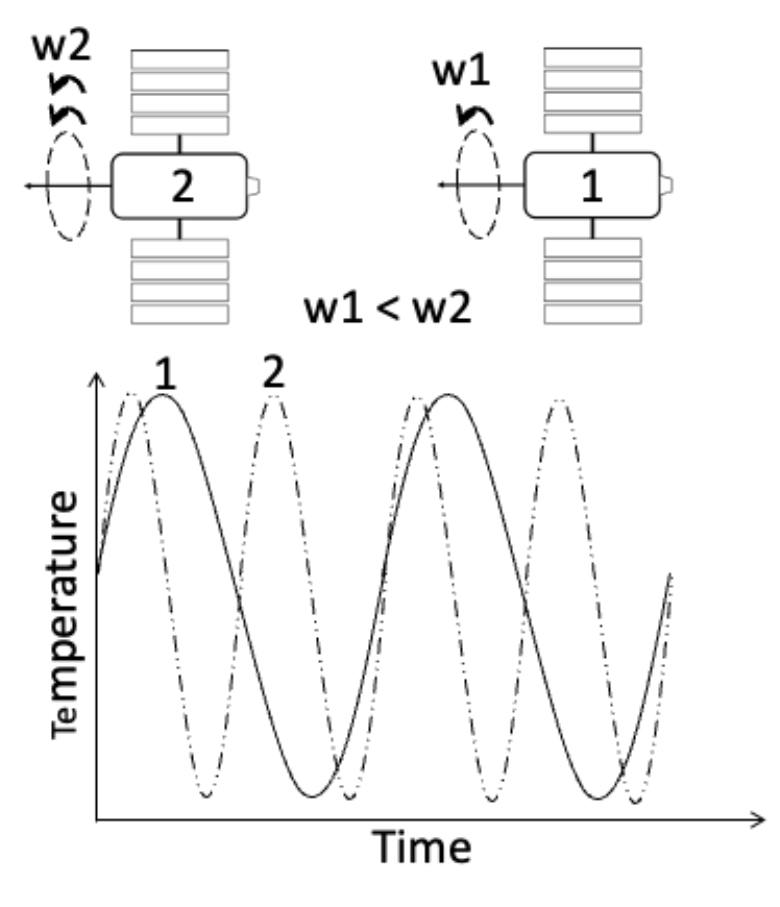}
	\caption{Effect of Rotational Motion with Different Angular Velocities ($\mathbf{w_1}$ and $\mathbf{w_2}$) on the Measurement of the CMB Temperature}
	\label{fig:4}
\end{figure}

Furthermore, the resulting temperature curve will resemble a sine wave if the spacecraft undergoes constant rotational motion (Figure \ref{fig:4}). In the figure, \(w_1\) and \(w_2\) represent the angular velocities of spacecraft $1$ and spacecraft $2$, respectively. Spacecraft $2$, with a higher angular velocity (\(w_2\)) compared to spacecraft 1 (\(w_1\)), exhibits a more rapid oscillation in the temperature curve. Introducing precession to this rotation results in a more complex yet cyclical curve behavior under steady rotation conditions.

Before delving into the velocity determination, an introduction to the scenario's conditions is essential. Consider a spacecraft in a circular orbit around a planet, positioned within an inertial coordinate system centered on the planet and aligned with the International Celestial Reference Frame (ICRF). This setup provides a stable reference frame for analyzing the spacecraft's motion and the effects of CMB measurements on velocity determination.

First, the component $T_F$ in Equation~(\ref{eq:14}) needs to be eliminated. This component can be considered Gaussian \cite{thommesen2020monte} and must be extracted from the sky signal. Other factors, such as systematic errors introduced during the calibration and filtering of raw sky data, are not considered here. Filtering techniques are crucial in data analysis for minimizing the impact of $T_F$, significantly improving raw data quality. This approach is instrumental in reducing noise and outliers, yielding a clearer and more accurate depiction of the underlying patterns or trends in the data. The primary aim is to distill essential information from the data while preserving its original structure.

For this research, the Savitzky-Golay method has been selected for filtering the CMB data due to its ability to reduce noise while maintaining the integrity of the signal's waveform. This characteristic is essential, as the waveforms are closely linked to the dynamics of spacecraft movement, as illustrated in Figures \ref{fig:2}, \ref{fig:3}, and \ref{fig:4}.

Initially developed for processing noisy data in chemical spectrometry, the Savitzky-Golay method \cite{Savitzky1964} excels at maintaining critical signal features such as peaks and valleys without distortion. Preserving these key signal characteristics is crucial for accurate data interpretation and application. This is especially significant in precise velocity estimation for spacecraft, where the integrity of signal features is paramount. Thus, the Savitzky-Golay method is optimal for the study's requirements, offering a balanced approach to noise reduction and feature preservation.

The Savitzky-Golay method relies on least-squares polynomial filtering. Generally, it involves replacing each point in a signal with a combination of signal values from a moving window centered around that point, based on the premise that adjacent points approximate the same fundamental value.

A notable challenge with this approach is the requirement for a batch of data to obtain an estimation due to the filtering process needing a data window. Consequently, this method does not facilitate real-time processing, resulting in sparser outputs. However, this is a minor issue in the context of IOD, as the primary impact is a delay in the information. This delay is acceptable given the nature of IOD, where immediate real-time processing is not critical, and the accuracy of the filtered data is of greater importance.

Furthermore, it is necessary to solve a nonlinear system to determine the magnitude and direction of the velocity. The complexity of solving this system depends on the prior knowledge of the velocity's magnitude or direction. This prior knowledge also impacts the number of sensors required. Consequently, two main approaches can be employed for velocity vector estimation, based on the availability of prior information:

\begin{itemize}
\item Without Prior Information: When the velocity's magnitude and direction are unknown, a nonlinear system can be constructed using a minimum of three different CMB readings at each time interval.
\item With Prior Information: When prior knowledge about the velocity's magnitude or direction is available through other navigation sensors, the number of sensors required to estimate the velocity can be reduced.
\end{itemize}

When prior knowledge regarding the magnitude and direction of velocity is lacking, a nonlinear system can be formulated by incorporating measurements from three CMB sensors at time $t$, $T_i(t): i=1,2,3$, and the unitary pointing directions of the sensors at time $t$, $\vec{n}_i=(a_i(t), b_i(t), c_i(t))$. This research considers the three sensors evenly distributed along a general direction. It is important to note that the velocity will be separated in magnitude over light speed, $v(t)/c$, and direction $(u_x(t), u_y(t), u_z(t))$ to make the variables dimensionless. Because of this, a fourth equation in the system will be necessary, which refers to the unitary magnitude of a direction vector. Thus, the representation of this nonlinear system \(\mathcal{S}\) at time $t$ is as follows (with $t$ omitted):

\begin{equation}
\label{eq:15}
\vec{n}_i=\left(a_i, b_i, c_i\right)
\end{equation}
\begin{equation}
\label{eq:16}
\vec{v}=v\left(u_x, u_y , u_z\right)
\end{equation}
\begin{equation}
\label{eq:17}
\mathcal{S}\left(v,\left[u_x, u_y, u_z\right]\right)=\left\{\begin{array}{c}
T_1-T_0 \frac{\sqrt{1-(v / c)^2}}{1-(v / c)\left(u_x, u_y, u_z\right) \cdot\left(a_1, b_1, c_1\right)} \\
T_2-T_0 \frac{\sqrt{1-(v / c)^2}}{1-(v / c)\left(u_x, u_y, u_z\right) \cdot\left(a_2, b_2, c_2\right)} \\
T_3-T_0 \frac{\sqrt{1-(v / c)^2}}{1-(v / c)\left(u_x, u_y, u_z\right) \cdot\left(a_3, b_3, c_3\right)} \\
u_x^2+u_y^2+u_z^2-1
\end{array}\right\}=0
\end{equation}

The velocity vector depicted in Equation~(\ref{eq:16}) does not directly represent the spacecraft's velocity relative to the planet. Instead, it results from various component vectors, which signify the velocity to the last scattering surface \cite{smoot2007nobel}. This distinction is crucial for understanding the spacecraft's motion within the broader cosmic context rather than merely its movement relative to a nearby celestial body. So, after solving the nonlinear system, Equation~(\ref{eq:17}), it is possible to isolate the spacecraft's velocity relative to the SSB as:
\begin{equation}
\label{eq:18}
\vec{v}_{s p}=\vec{v}-\vec{v}_{p}-\vec{v}_{S S B}
\end{equation}
where $\vec{v}_{p}$ is the planet's velocity relative to the SSB, obtained using ephemerides. The SSB's velocity, $\vec{v}_{SSB}$, is known to be $370$ km/s toward $(\alpha,\beta) = (264^\circ,48^\circ)$ \cite{Darling2022}, where $\alpha$ is the Galactic Longitude and $\beta$ is the Galactic Latitude. This represents the movement of the SSB relative to the last scattering surface.

The critical hurdle in deploying this strategy stems from addressing the nonlinear system, which calls for advanced mathematical strategies. Employing optimization techniques or iterative methods, particularly those inspired by the Newton method \cite{kelley2003solving}, is beneficial for navigating these nonlinear challenges. However, the particular conditions of this research add further intricacies to the resolution process. For instance, in calculations like those shown in Equation~(\ref{eq:17}), the $v/c$ ratio is crucial and might lead to the Jacobian matrix becoming singular. This ratio also threatens to significantly reduce numerical accuracy, especially as values near zero can alter substantially outcomes. Moreover, the value of this approach is highlighted when securing an accurate initial guess, which can be particularly challenging because of the lack of information.

In such instances, the Trust-Region-Dogleg Method \cite{Yingliang2000} becomes particularly valuable. This method retains its integrity even if the Jacobian matrix is singular and remains effective when the initial guess is far from the actual solution \cite{Powell1968}. However, the possibility of encountering local minima or other complex features typical of nonlinear systems necessitates meticulous attention and strategic management. Consequently, the selection and execution of the algorithm must be conducted with extensive testing and validation. 

When prior knowledge about the velocity's magnitude or direction is observed through other navigation sensors, reducing the number of sensors needed to estimate the velocity is possible. This simplifies the system to be solved and can enhance the precision of the results. The system to be solved in such scenarios at time $t$ is as follows:
\begin{itemize}
\item Velocity's direction known ($\hat{u}$):
\begin{equation}
\label{eq:19}
\mathcal{S}(v)=\left\{T_1-T_0 \frac{\sqrt{1-(v / c)^2}}{1-(v / c) \hat{u} \cdot\left(a_1, b_1, c_1\right)}\right\}=0
\end{equation}
\item Velocity's magnitude known ($v$):
\begin{equation}
\label{eq:20}
\mathcal{S}\left(u_x, u_y, u_z\right)=\left\{\begin{array}{c}
T_1-T_0 \frac{\sqrt{1-(v / c)^2}}{1-(v / c)\left(u_x, u_y, u_z\right) \cdot\left(a_1, b_1, c_1\right)} \\
T_2-T_0 \frac{\sqrt{1-(v / c)^2}}{1-(v / c)\left(u_x, u_y, u_z\right) \cdot\left(a_2, b_2, c_2\right)} \\
u_x^2+u_y^2+u_z^2-1
\end{array}\right\}=0
\end{equation}
\end{itemize}

One of the primary advantages of this last approach is the ability to leverage multiple sensors employing different technologies. This utilization of various sensors enhances the precision of the results and simplifies the handling of the nonlinear system by reducing the number of equations involved. However, the accuracy of this method heavily depends on the quality of the prior information. Any inaccuracies in the previous information could lead to significant errors in the velocity estimation. Therefore, it is essential to have reliable sources of previous information and account for possible uncertainties.

The focus of this research is solely on the system described in Equation~(\ref{eq:17}), as this provides a controlled environment for testing and validation. Additionally, focusing on a specific system allows for a more in-depth analysis and clearer presentation of the results.

\section{Velocity Estimation Using Machine Learning Regression Models}

Given the complexity of solving the non-linear system in Equation~(\ref{eq:17}), when prior knowledge about the velocity's magnitude or direction is unavailable, an alternative solution based on ML regression models is proposed.

The task of predicting the spacecraft's velocity using sensor pointing, sensor placement on the spacecraft, and CMB temperature measurements can be formulated as a regression problem. The dataset \( X \) with these input variables, can be described as follows:
\begin{equation}
\label{eq:21}
X = \{(\vec{n}_j(t), \vec{s}_j, T_j(t)) \mid j = 1, 2, \ldots, N\}
\end{equation}
where $\vec{n}_j(t)$ represents the pointing of the $j$-th sensor at time $t$ in the ICRF, $\vec{s}_j$ denotes the placement of the $j$-th sensor in the spacecraft frame, and $T_j(t)$ is the CMB temperature measurement from the $j$-th sensor at time $t$.

Given the input dataset $X$ from Equation (\ref{eq:21}) and the spacecraft's velocity vector at time $t$ ($\vec{v}_{sp}(t)$), which is the corresponding output variable, the objective is to find a function $f$ such that:

\begin{equation}
\label{eq:22}
\vec{v}_{sp}(t) = f(X) = f\left(\{(\vec{n}_j(t), \vec{s}_j, T_j(t)) \mid j = 1, 2, \ldots, N\}\right)
\end{equation}

This function $f$ should accurately model the relationship between the inputs and the output, minimizing prediction errors. The regression problem can thus be stated as finding the best fit for $f$ that maps the sensor pointing, sensor placements, and CMB temperature measurements to the spacecraft's velocity vector, with $N$ being the number of sensors. While Equation~(\ref{eq:17}) required three sensors, here $N$ can be any natural number. For this work, using $N=1$ is particularly attractive as it simplifies the system problem by reducing the number of sensors needed to estimate the spacecraft's velocity.

The aim is to minimize the error between the predicted velocity vector $\hat{\vec{v}}$ and the actual velocity vector $\vec{v}$ by optimizing the parameters of the function $f$. In practice, the model $f$ is trained using combined data from multiple sensors to capture the comprehensive relationship between the inputs and the spacecraft's velocity. Once trained, the model can be evaluated using individual sensor data to generate predictions, allowing for different results based on the specific sensor configuration.

In this format, the challenge becomes acquiring the right data and applying the adequate algorithm to find an efficient and accurate $f$. The data needs to faithfully represent the multiple readings of different spacecraft sensor configurations (sensor pointing and placement) in various orbit conditions. The model must be capable of predicting the current velocity with sufficient accuracy at a reasonable processing cost. The sufficient accuracy is determined by the limiting capability of the IOD method to handle \cite{christian2019initial}, and the processing cost will be related to the desired application.

\subsection{Regression Models}

Seeking alternative solutions to efficiently and accurately solve the regression problem, several ML-based models are evaluated, considering the advantages and disadvantages each one offers:

\begin{itemize}
    \item \textbf{Polynomial Regression} (PR): This technique extends a basic linear model by including higher-degree terms of the input variables and their interactions, allowing it to capture complex, non-linear relationships. The generated model is simple, a polynomial equation, and it can capture intricate patterns and relationships with higher degrees. Increasing the degree levels, however, rapidly increases complexity and can lead to overfitting and higher computational demand \cite{cheng2019polynomial};
    \item \textbf{Lasso and Ridge Regression}: These are regularized versions of PR that mitigate overfitting. Lasso regression (L1 regularization) uses a penalty proportional to the absolute values of the coefficients, while Ridge regression (L2 regularization) shrinks coefficients by adding a penalty proportional to their squared values. Lasso can perform feature selection by shrinking some coefficients to zero, making it useful for high-dimensional input spaces, while Ridge can be more efficient to fit in larger datasets \cite{Nelles2020linPollook};
    \item \textbf{Support Vector Regression} (SVR): This model handles non-linear relationships through kernel functions that project data into a higher-dimensional space for linear separation. It is resistant to outliers and provides robust generalization capabilities, at the cost of a more complex model compared to PR \cite{awad2015SVR};
    \item \textbf{Artificial Neural Networks} (ANN): ANNs are highly flexible and capable of modeling complex non-linear patterns. They are well-suited for processing large data volumes and learning intricate representations, though determining the ideal network structure can be challenging, and they generally lack interpretability \cite{soaresdoamaral2022metareview}; and
    \item \textbf{Random Forest} (RF): This solution utilizes an ensemble of decision trees to map inputs to outputs, proficiently handling non-linear relationships and offering resilience to overfitting. It can provide interpretable insights via feature importance scores but can be computationally demanding for extensive datasets and a high number of features \cite{biau2016random}.
\end{itemize}   

\subsection{Evaluation Metrics}

To compare the accuracy performance of regression alternatives, they are evaluated based on the Root Mean Squared Error (RMSE) and Mean Absolute Error (MAE) of the velocity predictions. MAE measures the average magnitude of the errors in a set of predictions, without considering their direction. It provides a straightforward interpretation of the average prediction error and is less sensitive to outliers. RMSE measures the square root of the average of the squared differences between predicted and actual values. This metric gives more weight to larger errors, making it particularly useful for highlighting significant deviations and penalizing models with substantial prediction errors. By incorporating both MAE and RMSE, we can comprehensively understand the model performance, balancing the need to understand average prediction accuracy and sensitivity to larger errors. For all experiments, the training process is repeated 30 times, with mean results with bootstrap\cite{davison1997bootstrap} 95\% confidence interval (CI) adopted as reference for statistical significance. 

In addition to accuracy metrics, the models' complexity and relative computational costs are compared based on two key factors: prediction time and the total number of learnable parameters. Prediction time refers to the duration required for the trained model to generate predictions on new data points and can vary depending on hardware and software implementations. For reference, the experiments were executed on a computer with a 12th Gen Intel(R) Core(TM) i9-12900KF 3.20 GHz processor and 64 GB of RAM, using the scikit-learn\cite{scikit-learn2011} library version 1.51 on a Windows 11 operating system. While this configuration is not directly comparable to potential spacecraft applications, this information is provided to facilitate a relative comparison among the models.

\subsection{Hyperparameter Fine-tuning}

Hyperparameter fine-tuning is a crucial step in the application and comparison of machine learning algorithms. It involves adjusting the key configuration settings of each algorithm to optimize its performance and ensure that it reaches its full potential after the training phase. The process of hyperparameter tuning aims to find the best combination of parameters that minimizes the model's prediction error, typically measured by RMSE. A range of values for the main hyperparameters was explored for each algorithm. 

For PR, the degree level is varied from 1 to 6. Both Lasso and Ridge regression models are fine-tuned with regularization coefficients ranging from \(1 \times 10^{-8}\) to 1.0 and degree levels from 1 to 6. SVR involves tuning the L2 regularization coefficient (from \(1 \times 10^{-8}\) to 1.0) and exploring different kernel functions (linear, polynomial, Radial Basis Function (RBF)) \cite{awad2015SVR}. For the ANN, the hyperparameters include the number of hidden layers (ranging from 2 to 6), nodes per layer (ranging from 16 to 256), L2 regularization coefficients (from \(1 \times 10^{-8}\) to 1.0), and various activation functions (relu, tanh, sinusoidal) \cite{Nelles2020linPollook}. Lastly, for the RF model, the number of estimators is adjusted from 10 to 200, and the maximum tree depth is varied from 20 to 200 \cite{biau2016random}.

\section{Experimental Simulation Scenario}

The scenario is chosen to be an orbit around the Earth; however, the simulation can be adapted for any other celestial body with minimal modifications. The orbital propagation is conducted using MATLAB, and an SGP4 propagator is utilized. The coordinate system employed is the Earth-Centered Inertial (ECI) frame, aligned with the ICRF.

\subsection{Scenario for Analytical Solution}

\begin{figure}[htb]
	\centering\includegraphics[width=3in]{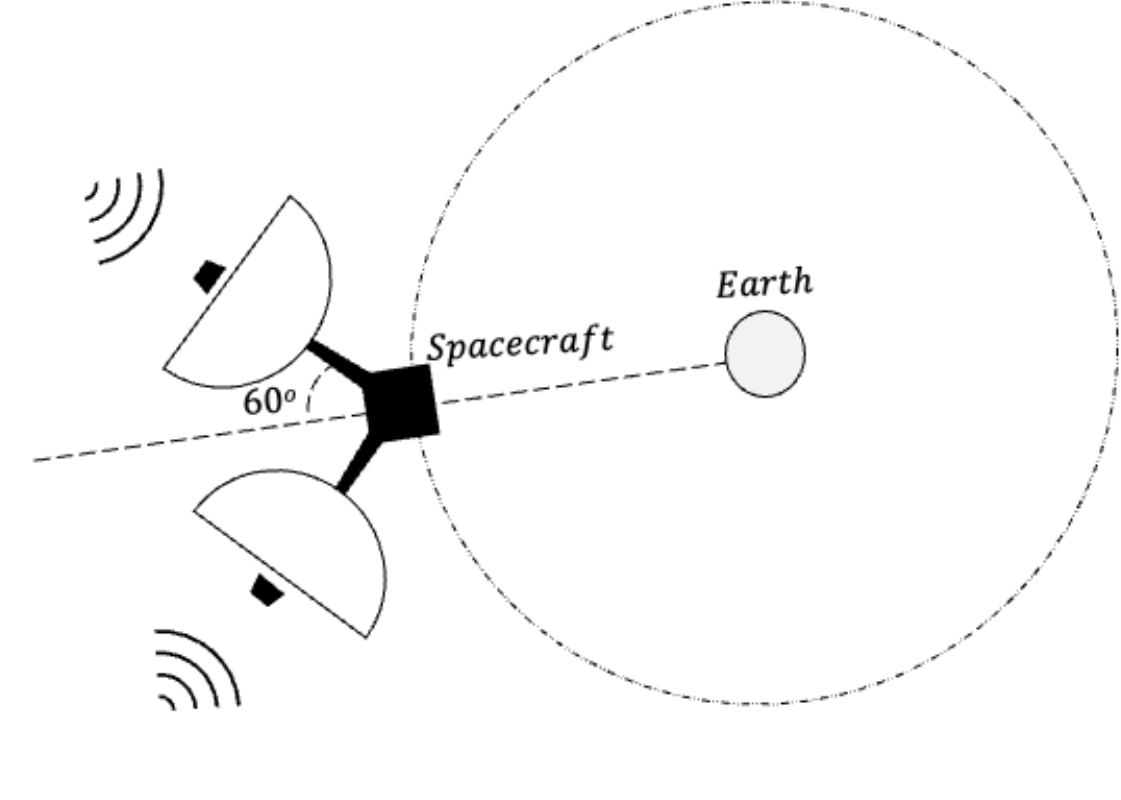}
	\caption{Schematics of the Orbit and the Sensors' Offset for the Simulation}
	\label{fig:5}
\end{figure}

Initially, for solving Equation~(\ref{eq:17}), three CMB sensors were simulated, positioned in the anti-direction of Earth, and equally distributed around this axis with an angular offset of $60$~degrees (Figure \ref{fig:5}), resulting in the sensors being spaced 120 degrees apart. The satellite follows an orbit around the Earth, maintained in a non-rotating state relative to a fixed body frame. The only necessary rotation is to sustain geopointing to the center of the Earth. Its altitude is set at $500$~km, with an inclination of $45$~degrees and an eccentricity of zero (circular orbit). The orbit was simulated over $6$~hours.

The CMB profile for the three sensors was generated using different values for $T_F$ ($0$~K, $10$~$\mu$K, $100$~$\mu$K, $150$~$\mu$K) in Equation~(\ref{eq:14}) to understand the influence of this parameter on the obtained velocity. $T_F$ varies with frequency and can be considered Gaussian; its value near $100$~GHz is approximately $100$~$\mu$K \cite{thommesen2020monte}.

\begin{figure}[htb]
	\centering\includegraphics[width=3.5in]{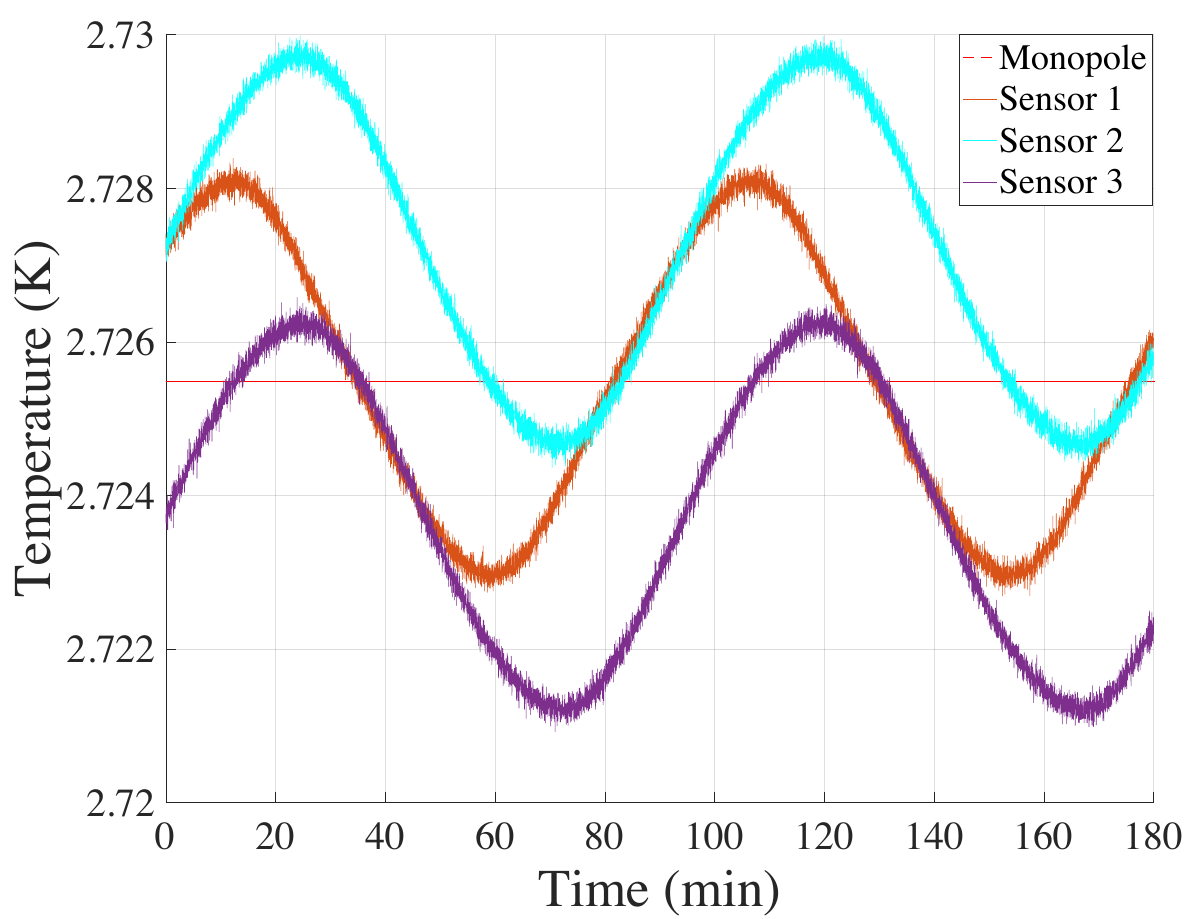}
	\caption{Simulation of CMB Measurement for the First 180 Minutes of Flight with $\mathbf{T_F=100}$~$\boldsymbol{\mu}$K}
	\label{fig:6}
\end{figure}

The changes in the CMB profile in this scenario are not primarily due to variations in the velocity magnitude. Although the magnitudes of the velocity vectors in Equation~(\ref{eq:18}) remain equal (circular orbit), their directions change, causing small variations in the velocity $\vec{v}$. These changes have a minimal impact on the results. The primary variations in the CMB temperature measurements are due to the changing angle between the sensor's pointing direction and the velocity vector, caused by the short orbital period. This is illustrated in Figure \ref{fig:6} for $T_F = 100$~$\mu$K.

\subsection{Scenario for ML Models Solution}

The scenario for evaluating the ML models is similar to the analytical solution, with a few key modifications to enhance the training and testing process. $T_F$ considered is just $100$~$\mu$K. Data was generated from 3,000 sensors with the same orbital elements and angular offset, but with different initial sensor pointing randomly distributed throughout the spacecraft. For each evaluation, 100 sensors were randomly selected for training, and 20 different sensors were selected for testing. From each selected sensor, 300 data points were randomly sampled from a total of 21,600 data points, these corresponding to a 6-hour flight with a data sampling frequency of 1 second. This process resulted in a training dataset consisting of 30,000 samples from 100 sensors and a testing dataset of 6,000 samples from 20 sensors.

Unlike the analytical scenario, where three sensor measurements are needed to estimate the velocity, the ML model can be built to require only one sensor measurement, i.e., $N=1$ in Equation~(\ref{eq:22}). This means that once the regression model is trained, it can generate accurate velocity predictions using any sensor positioned anywhere on the spacecraft. This capability significantly simplifies the implementation and improves the flexibility of the IOD process.

\section{Results and Discussions}

\subsection{Analytical Solutions Evaluation}

The CMB signal was filtered using the Savitzky-Golay method with a window of 1,500 samples and a polynomial order of 6. Assuming an initial velocity magnitude of $370$ km/s, which approximates the velocity of the SSB relative to the last scattering surface, and an initial null direction, the nonlinear system in Equation~(\ref{eq:17}) was solved using the Trust-Region-Dogleg Method.

The RMSE for different values of $T_F$ over 6 hours of flight is presented in Table~\ref{tab:analythical}. To calculate the RMSE, $5\%$ of the data from both the beginning and end were discarded due to potential instabilities caused by the Savitzky-Golay method in these regions.

\begin{table}[h]
\centering
\footnotesize
\caption{RMSE of Velocity for Each Component $\mathbf{(v_x,v_y,v_z)}$ at Different $\mathbf{T_F}$ Values}
\label{tab:analythical}
\begin{tabular}{llll}
\toprule
$T_F$~$(\mu\text{K})$ & X-RMSE (km/s) & Y-RMSE (km/s) & Z-RMSE (km/s) \\
\midrule
0 & 0.00 & 0.00 & 0.00 \\
5 & 0.04 & 0.04 & 0.04 \\
50 & 0.30 & 0.40 & 0.34 \\
100 & 0.64 & 0.72 & 0.56 \\
150 & 0.90 & 1.08 & 0.88 \\
\bottomrule
\end{tabular}
\end{table}

\begin{figure}[htb]
	\centering\includegraphics[width=3.5in]{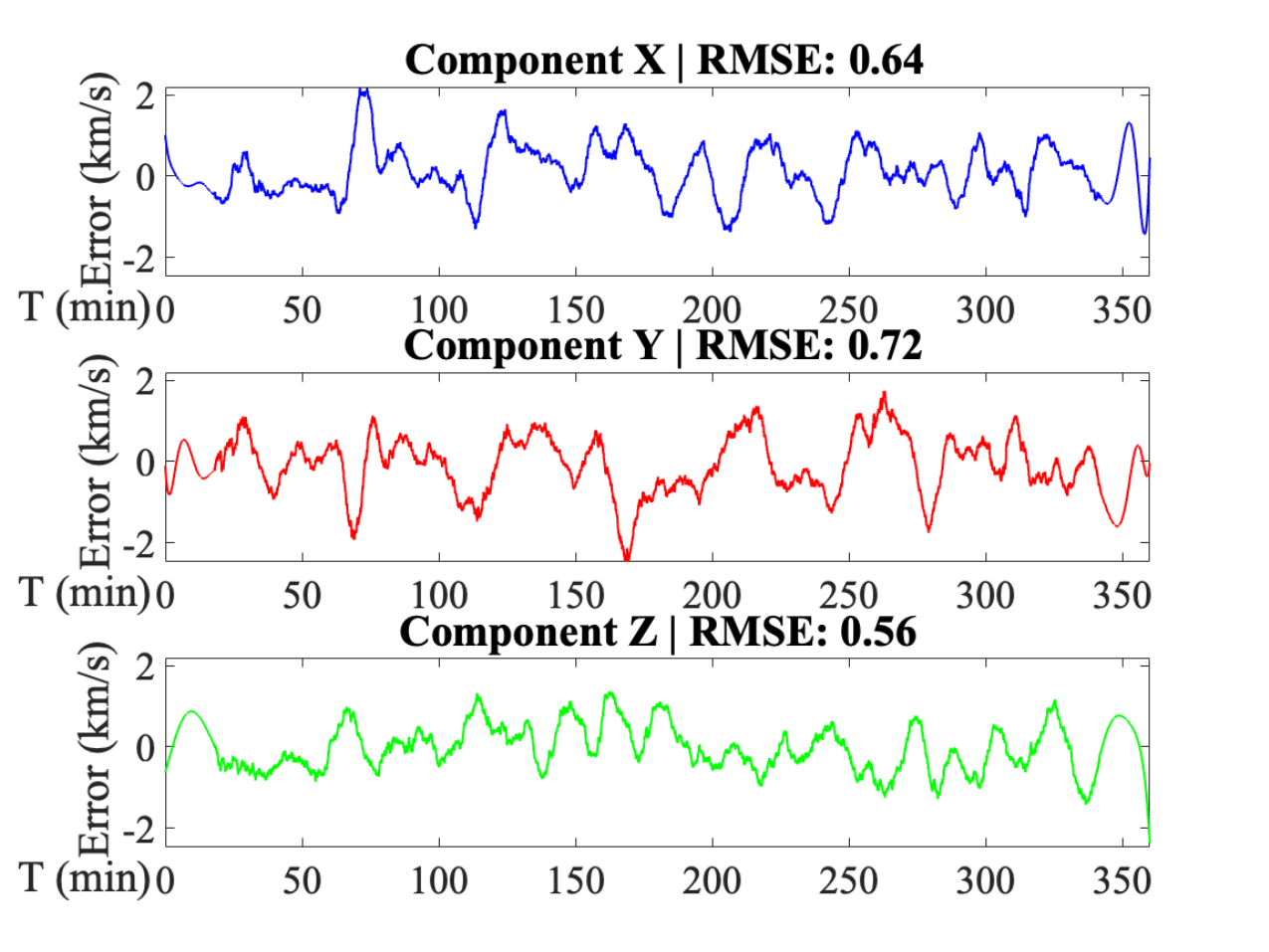}
	\caption{Velocity Estimation Error with 100~$\boldsymbol{\mu}$K Gaussian Noise over a 6-Hour Flight}
	\label{fig:7}
\end{figure}

The first two small values of $T_F$ are not realistic but serve to verify if the numerical solution approach introduces instabilities, which it does not. The two largest values are near-real conditions, and the errors for $100$~$\mu$K are illustrated in Figure \ref{fig:7}. A slight spike in the initial data point for all components can be observed, caused by the initial guess being somewhat distant from the true solution and the instabilities of the Savitzky-Golay method at the beginning. To mitigate this, the initial guess for each step was adjusted to the solution derived from the previous step.

As observed in Table~\ref{tab:analythical}, even with pre-filtering the CMB signal when $T_F > 50~\mu$K, the velocity estimation produces an RMSE greater than $0.3$ km/s. For errors around $10$~km in orbital altitude using velocity measurements equidistant in mean anomaly by about $100$~degrees, velocities with an RMSE below $0.01$ km/s are needed \cite{christian2019initial}. Therefore, new methods, such as ML models, are required for estimating velocity from the CMB in a more realistic scenario.

\subsection{Evaluation of ML Models }
The first step in evaluating the performance of the ML models involved optimizing the hyperparameters for each algorithm. Table \ref{tab:hyper} lists the selected parameters for evaluation, including specific settings, the regularization strategy, and the regularization coefficient values.

\begin{table}[htb]
\centering
\footnotesize
\caption{Model Parameters for the Evaluation of the CMB-Based Velocity Predictions}
\label{tab:hyper}
\begin{tabular}{llll}
\toprule
Model & Settings & Regularization & Coefficient \\
\midrule
PR & Degree level (6) & - & - \\
Lasso & Degree level (6) & L1 & $1 \times 10^{-4}$ \\
Ridge & Degree level (6) & L2 & $1 \times 10^{-7}$ \\
SVR & Kernel (RBF) & L2 & 0.002 \\
ANN & Layers (32, 32); Activation (Tanh) & L1 & 0.002 \\
RF & Estimators (30); MaxDepth (20) & - & - \\
\bottomrule
\end{tabular}
\end{table}

Considering the selected configurations, Table~\ref{Table_Final_Results} presents the final results for each algorithm. The best model overall, in terms of RMSE, was the Ridge regression model, which achieved an average RMSE of 0.0096 km/s. The PR model achieved very similar results with an RMSE of 0.0101 km/s, while Lasso did not perform at the same level, with an RMSE of 0.0178 km/s. For Ridge regression, the regularization significantly reduced the fitting time from 3.44 seconds for PR to 0.49 seconds. For Lasso, the regularization did not prove beneficial for either the results or the fitting time, which reached 3.99 seconds. The highest considered degree level of six yielded the best accuracy for the polynomial-based solutions; however, further investigations showed that increasing the level could potentially lead to marginally better results. The observed gains, however, were considered too marginal (on the order of $10^{-5}$) to justify the additional complexity and computational cost.

\begin{table*}[htb]
\centering
\footnotesize
\caption{Evaluation Results for Velocity Estimation Using Regression Models with Bootstrap 95\% CI}
\label{Table_Final_Results}
\begin{tabular}{l|cc|ccc}
\toprule
Model & MAE (CI) & RMSE (CI) & Fit Time (s) & Pred Time (ms) & Parameters \\
\midrule

PR & 0.0056 (0.0055, 0.0056) & 0.0101 (0.0099, 0.0102) & 3.44 & 65.0 & 3432 \\
Lasso & 0.0129 (0.0128, 0.0130) & 0.0178 (0.0178, 0.0179) & 3.99 & 58.0 & 3432 \\
Ridge & \textbf{0.0053 (0.0052, 0.0054)} & \textbf{0.0096 (0.0094, 0.0097)} & \textbf{0.49} & 59.0 & 3432 \\
SVR & 0.0464 (0.0462, 0.0464) & 0.0539 (0.0538, 0.0540) & 7.44 & 94.0 & \textbf{764} \\
ANN & 0.0247 (0.0245, 0.0252) & 0.0317 (0.0322, 0.0324) & 5.23 & \textbf{22.0} & 1411 \\
RF & 0.0293 (0.0291, 0.0294) & 0.0501 (0.0483, 0.0527) & 1.19 & 135.0 & 6650 \\
\bottomrule
\end{tabular}
\end{table*}

The ANN model exhibited the best prediction time performance, taking 22.0 ms compared to 59.0 ms or more for the polynomial-based solutions. Its RMSE of 0.0317 km/s, however, was 3.3 times higher than the Ridge result. The RF model performed worse, achieving a mean RMSE of 0.0501 km/s with a prediction time of 135.0 ms, similar to SVR with the RBF kernel, which achieved a mean RMSE of 0.0539 km/s and a prediction time of 94.0 ms.

\begin{figure}[htb]
\centering\includegraphics[width=3.5in]{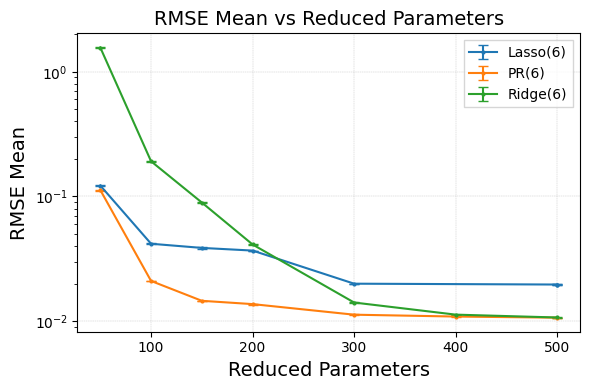}
\caption{Mean RMSE Results for Velocity Estimation Using Reduced Parameter Models from Polynomial-Based Algorithms}
\label{fig:8}
\end{figure}

An interesting observation is that the polynomial-based solutions were significantly slower in generating predictions compared to the ANN. This can be attributed to the high degree level of the polynomials used in these solutions. At the degree level of six, each polynomial-based solution (PR, Ridge, and Lasso) resulted in an equation with 3432 parameters, significantly more than the 1411 parameters in the ANN with two layers of 32 nodes each. The size of the polynomial solution, however, can be substantially reduced by eliminating the less significant factors, which are those with smaller absolute values in their coefficients. By doing so, the prediction performance of these models can be maintained even with a considerably smaller number of parameters. Figure \ref{fig:8} illustrates how reducing the number of parameters in the polynomial solutions impacts the RMSE. Notably, PR achieves better results than the ANN even with as few as 150 parameters, enabling predictions to be made in less than 1 ms. Furthermore, with 300 parameters, PR attains RMSE results equivalent to the complete model while reducing the prediction time to just 5 ms. Ridge and Lasso also allow for significant parameter reduction. Lasso achieves equivalent results to the complete model with 300 parameters but still with lower accuracy. Ridge shows the worst results with fewer parameters but achieves the complete model results with 500.

These results suggest that Ridge and PR models can be robust solutions for predicting spacecraft velocity in orbit based on CMB temperature readings, with polynomial reduced models achieving the best accuracies with lower computational costs. The PR-based model's best prediction accuracy and reduced computational cost after reduction make it an attractive option despite not being commonly preferred for this type of problem \cite{soaresdoamaral2022metareview}. This superior performance can be attributed to the full utilization of the polynomial solution, generating all interaction factors even at higher degree levels, which is often neglected in many regression evaluations \cite{cheng2019polynomial}. The simplicity of the final PR-based models, being just a polynomial equation, also makes them easily transferable to onboard systems.

Given these results, the Ridge model reduced to 500 parameters, with prediction errors below $0.01$~km/s, was selected as the reference for evaluating the practical application of the regression model for IOD.

\subsection{IOD Evaluation Using Ridge Model}

To test the 6th-order Ridge model with $500$ parameters, $50$ sensors with initial pointing not used in the model's training process were generated. It is important to note that estimating velocity from CMB signals using ML models requires only one sensor reading, compared to the three required in the analytical approach. This is a significant advantage of using this method.

\begin{figure}[htb]
	\centering\includegraphics[width=3.5in]{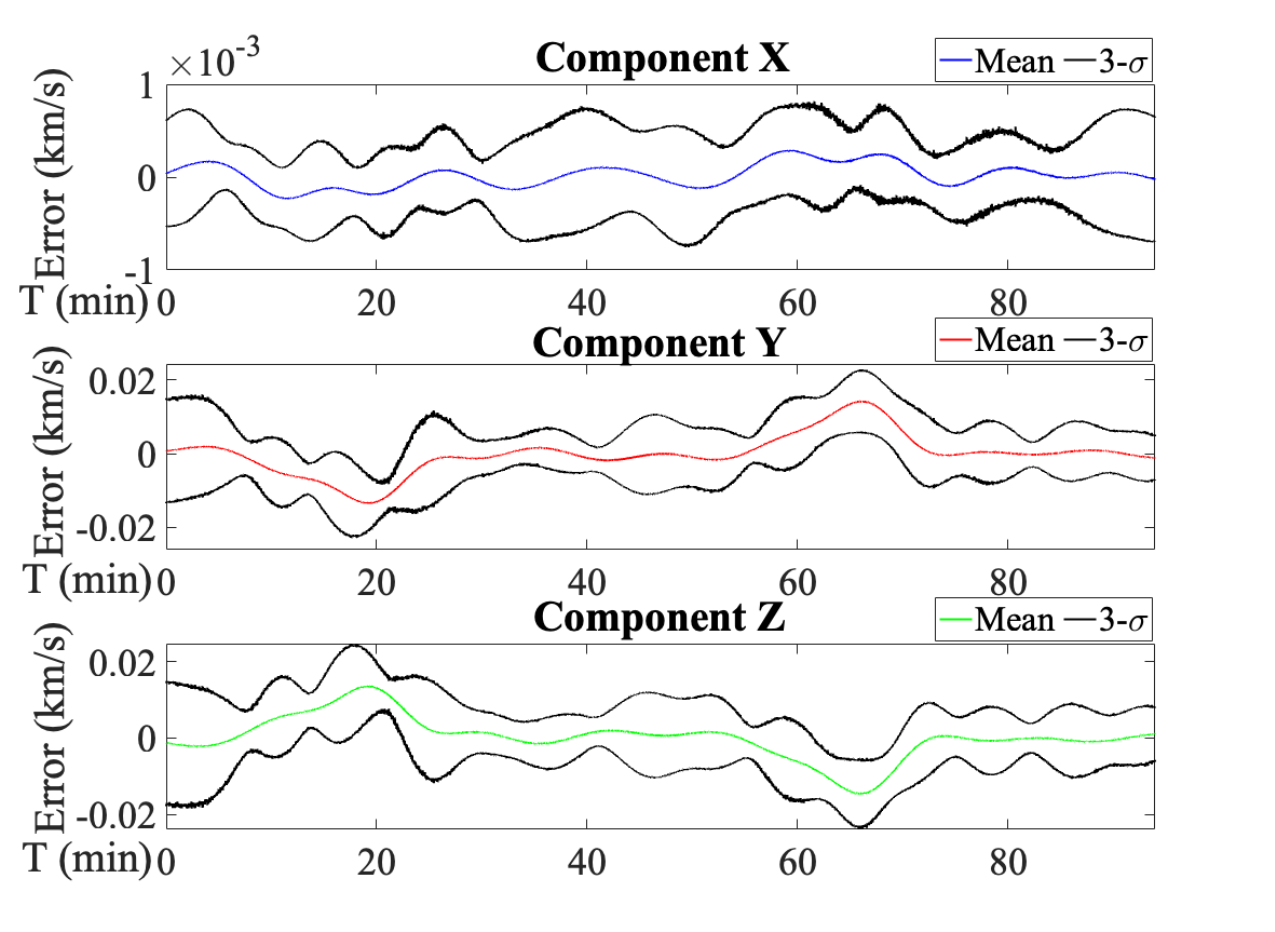}
	\caption{Velocity Estimation Mean Error with 3-$\boldsymbol{\sigma}$ for Ridge Model Applied to 50 Sensors with 100~$\boldsymbol{\mu}$K Gaussian Noise for 1 Orbit at 500 km and $\mathbf{45^o}$ Inclination}
	\label{fig:9}
\end{figure}

The results for the velocity estimation error over the duration of one orbit, depicted in Figure \ref{fig:9}, include a $3$-$\sigma$ curve. It can be seen that the errors for the x-component are much smaller due to the simplification process using this component as a reference. The mean value for all components is less than $0.01$~km/s. However, some individual values are greater than $0.01$~km/s but less than $0.02$~km/s. Consequently, it is expected that the IOD error fluctuates at values greater than $10$~km \cite{christian2019initial}.

\begin{figure}[htb]
	\centering\includegraphics[width=3.5in]{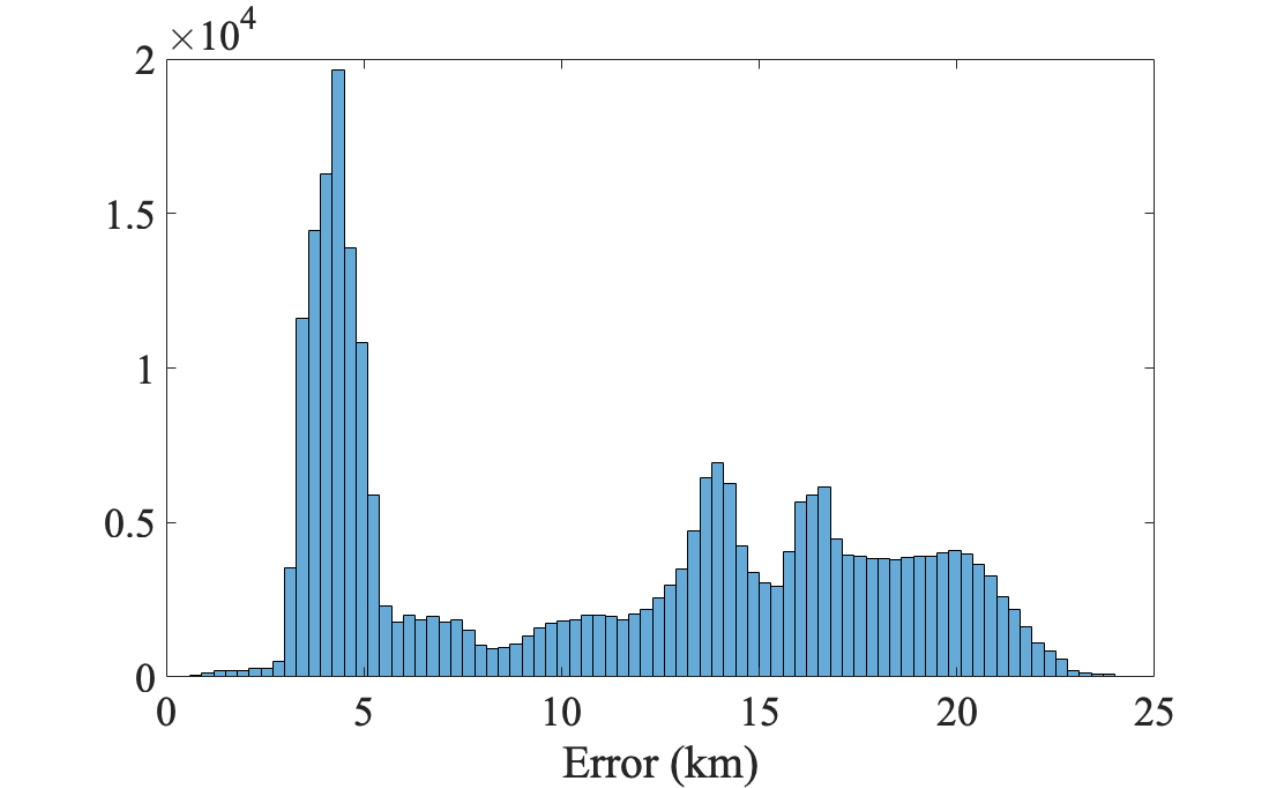}
	\caption{Histogram of Position Estimation Errors for Ridge Model Applied to $\mathbf{50}$ Sensors with $\mathbf{T_F=100}$~$\boldsymbol {\mu}$K over 1 Orbit at an Altitude of 500 km and an Inclination of $\mathbf{45^\circ}$}
	\label{fig:10}
\end{figure}

For the IOD, a triplet of estimated velocities, equally spaced at $120$ degrees in mean anomaly, is used. This approach generates a corresponding triplet of positions after applying the process described in Equations~(\ref{eq:1})-(\ref{eq:12}). A total of $1,800$ different triplets of estimated velocities are selected from each of the $50$ sensors' results, resulting in a population of $270,000$ data points. The histogram of the norm of the position vector error, calculated from the difference between the actual and estimated vectors, is shown in Figure \ref{fig:10}.

The modal interval was $4.2-4.5$~km in position estimation error. Due to some velocity estimation errors being greater than $0.01$~km/s, some position errors ranged between $14-20$~km. The mean position error was $11.12$~km, aligned with expectations from the literature for this level of velocity error \cite{christian2019initial}.

Thus, the Ridge method demonstrates reasonable results and has three significant advantages over analytical methods. First, it requires only one sensor reading. Second, the velocity estimation process achieves better results than the analytical approach for realistic values of $T_F$. Third, pre-filtering the CMB signal was unnecessary as the ML model could handle the noise independently. However, a disadvantage is that if the actual scenario deviates far from the training data, the errors can be much higher. Another disadvantage is the training process itself, which can be tricky and requires several tests to optimize the hyperparameters.

\FloatBarrier

\section{Conclusion}

This research successfully demonstrates the innovative use of CMB radiation as a reference signal for IOD, offering a groundbreaking method for estimating spacecraft velocity and position with minimal reliance on pre-existing environmental data. This approach, leveraging ML regression models, particularly the 6th order Ridge model with 500 parameters, has proven highly effective in determining velocity from CMB signals and, subsequently, the satellite's position using the three velocity vectors method.

The results underscore CMB's potential to enhance spacecraft operations' autonomy and flexibility, making it a valuable asset for space missions that do not rely on Earth-specific conditions or extensive ground-based infrastructure. However, the research also acknowledges certain limitations, such as the technological capability to measure the CMB's slight variations due to spacecraft motion, the impact of implementing CMB sensors on spacecraft platforms, potential high errors if the actual scenario deviates significantly from the training data, and the complexity of the training process, which requires meticulous hyperparameter optimization.

Future research could explore several avenues to further enhance and validate the use of CMB radiation for IOD. One area of focus is developing more sensitive and precise CMB sensors capable of accurately measuring the slight variations in CMB radiation caused by spacecraft motion. Additionally, investigating and developing more robust ML models is essential to handle a wider range of scenarios and reduce the potential for high errors when actual conditions deviate from the training data.

%%%%%%%%%%%%%%%%%%%%%%%%%%%%%%%%%%%%%%%%%%%%%

\bibliographystyle{AAS_publication}   % Number the references.
%\bibliography{references}   % Use references.bib to resolve the labels.

\end{document}